\documentclass[aps,prl,floatfix,twocolumn,amsfonts,amssymb,amsmath,superscriptaddress]{revtex4-1} 
\usepackage{dsfont} 
\usepackage{float} 
\usepackage[colorlinks=true,linkcolor=black,citecolor=black,filecolor=black,urlcolor=black,breaklinks=true]{hyperref}
\usepackage{mathtools}
\usepackage[capitalize]{cleveref}
\usepackage{physics}
\usepackage{tikz}
\usepackage{ulem} 
\usepackage{xcolor}
\usepackage{csquotes}
\usepackage{ifthen}

\newcommand{\avg}[1]{\left \langle #1 \right \rangle}

\DeclarePairedDelimiterX{\infdivx}[2]{(}{)}{%
#1\;\delimsize\|\;#2%
}

\newcommand{\aff}[1][default]{
  \ifthenelse{\equal{#1}{default}}{
    \boldsymbol{\lambda}
  }{
  \boldsymbol{\lambda^{#1}}
  }
}

\DeclarePairedDelimiterX\inproduct[2]{\langle\negthinspace\negthinspace\langle}{\rangle\negthinspace\negthinspace\rangle}{#1 \delimsize\vert #2}

\DeclarePairedDelimiterX\superdyad[2]{\lvert}{\rvert}{#1 \delimsize\rangle\negthinspace\negthinspace\rangle \negthinspace\langle\negthinspace\negthinspace\langle #2}

\begin{document}

\title{Far-from-equilibrium thermodynamics of non-Abelian thermal states}
\author{Franklin L. S. Rodrigues}
\affiliation{Institute for Theoretical Physics I, University of Stuttgart, D-70550 Stuttgart, Germany}
\affiliation{Physikalisches Institut, Albert-Ludwigs-Universität Freiburg, Hermann-Herder-Straße 3, D-79104 Freiburg, Germany}
\affiliation{EUCOR Centre for Quantum Science and Quantum Computing, Albert-Ludwigs-Universität Freiburg, Hermann-Herder-Straße 3, D-79104 Freiburg, Germany}
\author{Eric Lutz}
\affiliation{Institute for Theoretical Physics I, University of Stuttgart, D-70550 Stuttgart, Germany}

\begin{abstract}Noncommutativity of observables is a central feature of quantum physics. It plays a fundamental role in the formulation of the uncertainty principle for complementary variables and strongly affects the laws of thermodynamics for systems with noncommuting, that is, non-Abelian, conserved quantities. We here derive nonequilibrium generalizations of the second law of  thermodynamics in the form of fluctuation relations, both for mechanically and thermally driven quantum systems. We identify a non-Abelian contribution to the energy and entropy balances, without which  these relations would be violated. The latter term can be controlled to enhance both work extraction and nonequilibrium currents compared to what is obtained in commuting thermodynamics. These findings demonstrate that noncommutativity maybe a useful thermodynamic resource.
    \end{abstract}

\maketitle 

Statistical mechanics offers a generic  probabilistic framework to describe the equilibrium thermodynamics of systems with conserved quantities. In this theory, average macroscopic properties  are determined from the probability distribution (often called Gibbs distribution) of the corresponding microscopic states \cite{cal85}. Such  distributions may be obtained by maximizing the information entropy of the system with the  constraints of normalization and average conserved quantities \cite{jay57,jay57a,bal86,bal87}. A prominent example is  provided by the grand canonical distribution, which is associated with  conserved average energy and  average particle number \cite{cal85}. More generally, the powerful maximum entropy approach has been extended to an arbitrary number of conserved quantities \cite{jay57,jay57a,bal86,bal87}, leading to the concept of generalized Gibbs ensembles, which play a central role in  the description of relaxation dynamics in isolated integrable quantum systems \cite{vid16}. 

Whereas conserved quantities always commute in classical statistical physics, this is not necessarily the case in the quantum domain \cite{yun16,gur16,los17,pop20,fuk20,yun20,man22,mad23,upa24}. Owing to their noncommutation, conserved quantum observables  do not in general have well-defined simultaneous values. The resulting  thermal states  are  frequently referred to as non-Abelian thermal states to emphasize  their  noncommutative feature \cite{yun16,gur16,los17,pop20,fuk20,yun20,man22,mad23,upa24}. They have been shown to  exhibit unconventional nonclassical features that differ from those of common thermal states \cite{yun16,gur16,los17,pop20,fuk20,yun20,man22,mad23,upa24}, such as the absence of microcanonical subspaces \cite{yun16}. The experimental observation of the nonstandard thermalization towards a non-Abelian thermal state has recently been reported  for a  long-range Heisenberg chain in an ion-trap setup \cite{fra22}.

Nonequilibrium properties of small quantum systems are dominated by thermal and  quantum  fluctuations  \cite{esp09,cam11}. The second law  needs therefore to be generalized to account for fluctuating thermodynamic variables. One important such stochastic extension is provided by fluctuation relations of the form, $P(\Sigma)/P(-\Sigma) = \exp(\Sigma)$, for the random  entropy production $\Sigma$ with probability distribution $P(\Sigma)$ \cite{esp09,cam11,def11,lan21}. Fluctuation theorems imply the second law on average, $\langle \Sigma \rangle \geq 0$, and quantify the occurrence of negative entropy production events. They are moreover valid arbitrarily far from equilibrium. They have, for this reason, found widespread applications in the study of nonequilibrium microscopic systems \cite{esp09,cam11}. One usually distinguishes work fluctuation relations, with entropy production $\Sigma = \beta (W - \Delta F)$, for systems that are mechanically driven from a thermal state at inverse temperature $\beta$ ($\Delta F$ here denotes the associated free energy difference) \cite{jar97} and heat exchange fluctuation relations, with entropy production $\Sigma =\Delta \beta Q$, for two thermally coupled systems with  inverse temperature difference $\Delta \beta$ \cite{jar04}. Fluctuation theorems  have lately been obtained for generalized Gibbs ensembles with commuting conserved quantities \cite{hic14,mur18,mur20}. However, general quantum fluctuation relations for systems with noncommuting conserved quantities are currently missing. 

We here investigate the far-from-equilibrium thermodynamics of generalized Gibbs ensembles with noncommuting conserved quantities   by deriving  both work and exchange quantum fluctuation theorems for non-Abelian thermal states. Since there is no joint eigenbasis for noncommuting observables, off-diagonal density matrix elements, associated with quantum coherence \cite{str17},  appear in the   eigenbasis of the conserved observables.   Such quantum coherence strongly impacts the nonequilibrium properties of non-Abelian thermal states. However, standard methods, such as the two-point-measurement scheme, which has been extensively used to study quantum fluctuations relations  \cite{esp09,cam11}, are not applicable in this case. Owing to their projective nature, they indeed cannot account for nondiagonal matrix elements of initial or final states of a  quantum process \cite{esp09,cam11}. In the following, we   instead employ  the formalism of dynamic Bayesian networks   that  allows one to analyze conditional dependencies in a general set of time-dependent random variables via Bayes' rule \cite{nea03,dar09}. As a result, this  approach is able to capture off-diagonal density matrix elements at all times \cite{mic20,par20,str20,mic21,rod24,mic24}. We identify, in particular, a new non-Abelian contribution to the first and second laws, and show that the latter can be exploited to enhance both work extraction and exchange currents, as well as their  thermodynamic efficiency, defined by their respective ratio to the nonequilibrium entropy production.

\textit{Generalized Gibbs ensemble.}  We consider a system initialized in a generalized Gibbs state with an arbitrary  number $N$ of (possibly noncommuting) conserved quantities $A_k$, also called charges,  and corresponding affinities $ \lambda_k$. Its   density operator is given by \cite{yun16,gur16,los17,pop20,fuk20,yun20,man22,mad23,upa24}
\begin{equation}
\label{1}
\rho = \exp(F - \boldsymbol{\lambda} \cdot \boldsymbol{A}),
\end{equation}
where, to simplify notation,  we have introduced the  vectors $\boldsymbol{\lambda}$ and $ \boldsymbol{A}$, with respective components $\lambda_k$ and $A_k$, such that  $\boldsymbol{\lambda} \cdot \boldsymbol{A} = \sum_k \lambda_k A_k$, as well as the generalized free entropy  $F = -\ln\Tr\big[\exp(-\boldsymbol{\lambda} \cdot \boldsymbol{A})]$, also known as the Massieu potential \cite{cal85}. Equation \eqref{1} generalizes the usual canonical distributions of statistical mechanics \cite{cal85}. One of the observables, say $A_1$, might be the energy of the system, in which case $\lambda_1$ is the inverse temperature, but the energy need not be one of the conserved quantities  \cite{yun16,gur16,los17,pop20,fuk20,yun20,man22,mad23,upa24}. For the standard grand canonical distribution, the second affinity is the product of inverse temperature and chemical potential \cite{cal85}. Since the observables $A_k$ do not necessarily commute, $[A_k,A_l\big]\neq0$,  eigenbases associated with different charges  are  not always  mutually orthogonal, contrary to the case of commuting observables; in the grand canonical ensemble, for instance, energy and particle number operators  alway commute, and therefore have a common eigenbasis \cite{cal85}. 

\textit{First law  for noncommuting charges}. Let us begin by deriving a generalized first law of thermodynamics for non-Abelian states. To that end, we drive the system, initially prepared in state \eqref{1} at time $t=0$, via a nonequilibrium protocol of duration $\tau$, parametrized by  time-dependent  charges $\boldsymbol{A}^t$. During the entire protocol, the system is weakly coupled to a non-Abelian reservoir,  $\rho^R = \exp(F^R - \boldsymbol{\lambda}^R \cdot \boldsymbol{A}^R)$, through an interaction Hamiltonian $V$. For concreteness, we  write the interaction in the general form, $V = \sum_\alpha S_\alpha \otimes R_\alpha$, as usually done \cite{bre02}, where $S_\alpha$ and $R_\alpha$ are respective system and reservoir operators. We further take it to satisfy strict energy conservation at all times, $[V, \boldsymbol{A}^t + \boldsymbol{A}^R] = 0$ \cite{cic22}, as, for example, for a rotating-wave coupling in quantum optics. We additionally denote  the projectors on the respective instantaneous eigenstates of system and bath as $\Pi_i^t$ and $\Pi^R_\mu$, and express the corresponding non-Abelian states as $\rho^t = \sum_i p^t_i\Pi_i^t$ and $\rho^R = \sum_\mu p^R_\mu \Pi^R_\mu$, with associated probabilities $p_i^t$ and $p^R_\mu$; we will use latin (greek) indices for system (reservoir) operators throughout. 

A stochastic first law  for a single realization of the nonequilibrium quantum process can then be obtained in the presence of noncommuting charges by introducing the local charge change $\Delta \boldsymbol{a}(i^0,j^\tau) = \mel{j}{\boldsymbol{A}^\tau}{j} - \mel{i}{\boldsymbol{A}^0}{i}$ and the heat  $\boldsymbol{q}(\mu,\nu) = -(\mel{\nu}{\boldsymbol{A}^R}{\nu} - \mel{\mu}{\boldsymbol{A}^R}{\mu})$ exchanged with the bath, in analogy with  the standard derivation of the first law for thermal Gibbs states \cite{esp09,cam11}. We find
\begin{equation}\label{2}
    \Delta \boldsymbol{a}(i^0,j^\tau) =  \boldsymbol{w}(i^0,\mu,j^\tau,\nu) + \boldsymbol{q}(\mu,\nu) + \boldsymbol{\varepsilon}(i^0,\mu,j^\tau,\nu),
\end{equation}
with the stochastic work $\boldsymbol{w}(i_0,\mu,j_\tau,\nu)$ and a new non-Abelian contribution (Supplemental Material)
\begin{eqnarray}\label{3}
  \boldsymbol{\varepsilon}(i^0,j^\tau,\mu,\nu) =&& \frac{\mel{j^\tau, \nu}{\boldsymbol{\mathcal{O}^\tau} V - V \boldsymbol{\mathcal{O}^0}}{i^0, \mu}}{\mel{j^\tau, \nu}{V}{i^0, \mu}} \nonumber\\
  && -\frac{\mel{j^\tau}{\boldsymbol{\Delta A}}{j^\tau} - \mel{i^0}{\boldsymbol{\Delta A}}{i^0}}{2},
\end{eqnarray}
where we have defined $\boldsymbol{\Delta A} = \boldsymbol{A^\tau} - \boldsymbol{A^0}$ and the operators $\boldsymbol{\mathcal{O}^\tau} = (\boldsymbol{A^\tau} (\mathds{1} - \Pi^\tau_j) + (\mathds{1} - \Pi^\tau_j) \boldsymbol{A^0})/2 + \boldsymbol{A^R}(\mathds{1} - \Pi^R_\nu)$ and $\boldsymbol{\mathcal{O}^0} = (\boldsymbol{A^\tau}(\mathds{1} - \Pi^0_i) + (\mathds{1} - \Pi^0_i)\boldsymbol{A^0})/2 + (\mathds{1} - \Pi^R_\mu) \boldsymbol{A^R}$. Expression \eqref{3} vanishes for commuting charges, and is hence absent in the standard formulation of the energy balance. It may be interpreted as the work done by the non-Abelian reservoir on the system, in addition to that performed by the external time-dependent driving, extending this notion known from the coupling to squeezed environments \cite{nie16,nie18,man18} to arbitrary non-Abelian baths.

\begin{figure*}[t]
	\centering
	\begin{tikzpicture}
	\node (a) [label={[label distance=-.4 cm]145: \textbf{a)}}] at (0,0) {\includegraphics[width=0.32\textwidth]{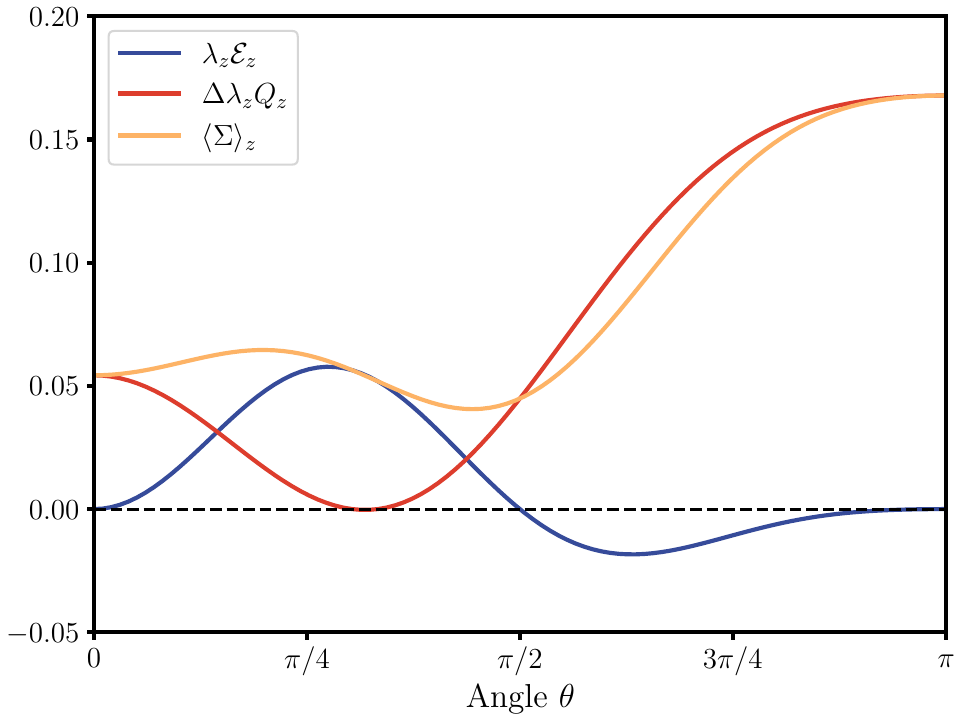}};	\node (a) [label={[label distance=-.4 cm]145: \textbf{b)}}] at (6,0) {\includegraphics[width=0.32\textwidth]{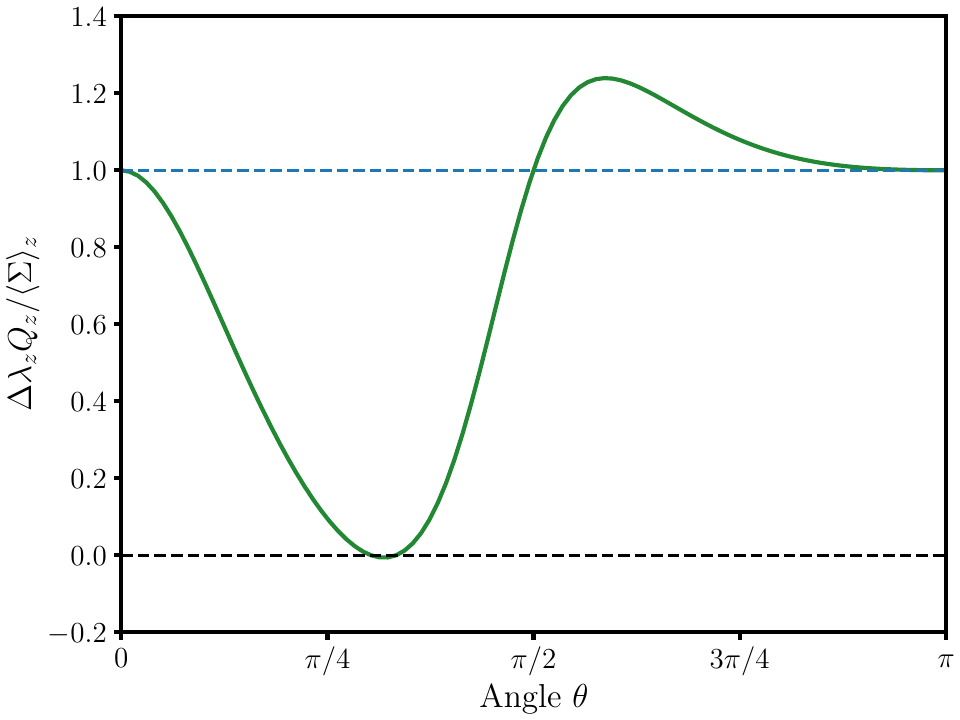}};
	\node (a) [label={[label distance=-.4 cm]145: \textbf{c)}}] at (12.0,0) {\includegraphics[width=0.32\textwidth]{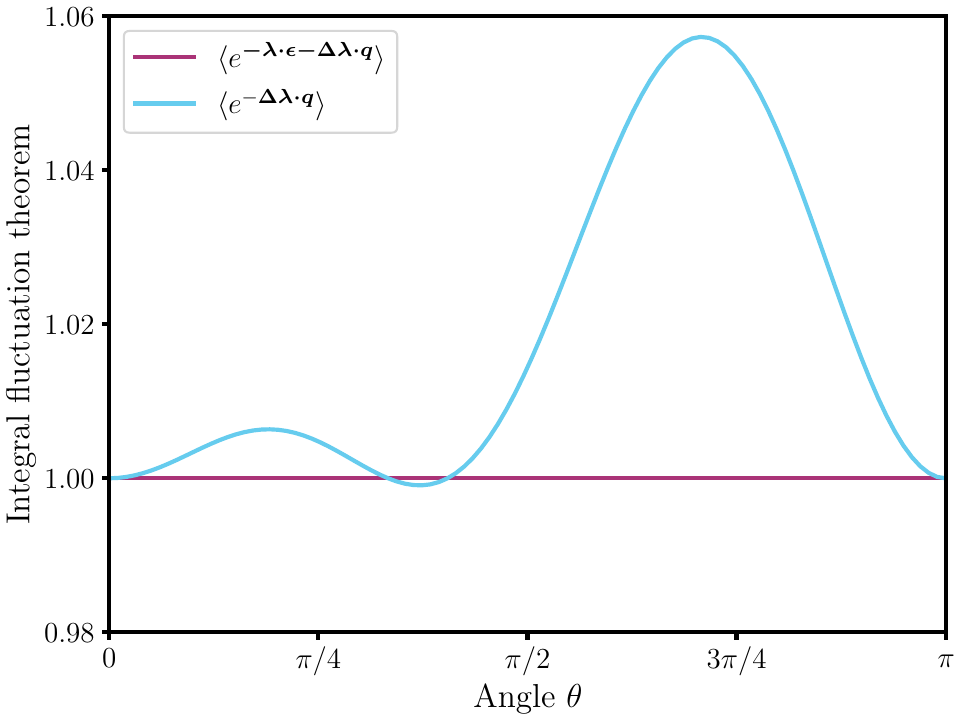}};
	\end{tikzpicture}
	\caption{Non-Abelian exchange fluctuation relation. a)  Non-Abelian contribution ${\mathcal{E}}_z$,  current $ {Q}_z$ and  entropy production $ \avg{\Sigma}_z$, for a two-qubit Heisenberg chain, as a function of the angle $\theta$ between the spin states. In the commuting case, $\theta=(0, \pi)$, ${\mathcal{E}}_z$ vanishes, implying that entropy production and nonequilibrium current are equal. For $\pi/2<\theta<\pi$, the non-Abelian term is negative. b) In this noncommuting regime, the ratio of current and dissipation is enhanced compared to the Abelian case, showing that noncommuting charges are a thermodynamic resource. c) The exchange fluctuation relation (11) is only satisfied, when the non-Abelian contribution ${\boldsymbol{\mathcal{E}}}$, Eq.~(3), is included.    Parameters are $J = \omega = 1$, $\beta = 1$, $\beta^R = 0.5$ and $\tau = \pi$.}\label{fig1}
\end{figure*}

\textit{Non-Abelian work fluctuation relation}.  We next derive an extension of the second law in the form of a Jarzynski-Crooks-type fluctuation relation \cite{esp09,cam11}.  As the initial system state is in general  coherent in the environmental eigenbasis, the usual two-point measurement scheme  \cite{esp09,cam11} is inappropriate to study the fluctuations of the protocol $\boldsymbol{A}^t$, since it destroys nondiagonal density matrix elements, owing to its  projective nature \cite{tal07}. We address this issue by  employing the tools of dynamic Bayesian networks, which allow one to account for off-diagonal matrix elements at all times, by specifying  the dynamics of a system in an eigenbasis conditioned on the evolution in an incompatible eigenbasis   \cite{mic20,par20,str20,mic21,rod24,mic24}.
Dynamic Bayesian network are commonly used to  describe the  relationship between dynamical variables through conditional probabilities evaluated via Bayes' rule \cite{nea03,dar09}; they may be viewed as an extension of hidden Markov models.

The conditional probability of measuring the system in state $\ket{m^t}$ at time $t$ given that it is in state $\ket{i^t}$ is given by Born's law, $p^t(m\vert i)=\abs{\braket{m^t}{i^t}}^2$; these constitute the conditional prior probabilities, which are to be determined beforehand. If the environment is thermal, this amounts to energy measurements on the system. By performing conditional measurements at the beginning and at the end of the nonequilibrium  driving protocol, we may  introduce a conditional trajectory $\Gamma = (i,\mu,m,j,\nu,n)$ for the composite system, with path probability \cite{mic20,par20,str20,mic21,rod24,mic24}
\begin{equation}
  \begin{split}
    P(\Gamma) = p^0_i p^R_\mu p^0(m \vert i) p^{0,\tau}(j, \nu \vert i, \mu) p^\tau(n\vert j),
  \end{split}
  \label{4}
\end{equation}
where $p^{0,\tau}(j \vert i) = \abs{\mel*{j^\tau, \nu}{U}{i^0, \mu}}^2$ is the (transition) probability  of measuring state $\ket{j^\tau}$ after evolving state $\ket{i^0}$ through the joint system-environment time evolution operator $U = U(\tau,0)$. We emphasize  that the measurement  at the end of the protocol is performed in the eigenbasis  of the instantaneous generalized Gibbs state $\pi^t = \exp(F - \boldsymbol{\lambda} \cdot \boldsymbol{A}^t)$  (and not in instantaneous eigenbasis of $\rho^t$)  in accordance to the original spirit of the Jarzynski-Crooks relation \cite{cro99}.

We may similarly define the probability of the reversed trajectory $\Gamma^\dagger = (j,\nu, n, i,\mu, m)$ as
\begin{equation}
\label{5}
  \begin{split}
    P^\dagger(\Gamma^\dagger) = p^\tau_i p^R(\nu) p^\tau(n \vert j) p^{\tau,0}(i, \mu \vert j, \nu) p^0(m \vert i),
  \end{split}
\end{equation}
where the initial state of the reservoir is chosen as the final instantaneous generalized Gibbs state of the system, $\rho^{R,0} = \pi^\tau$, and where $p^{\tau,0}(i, \mu \vert j, \nu) = \abs{\mel*{i^0, \mu}{U^\dagger}{j^\tau, \nu}}^2$, with the time-reversed evolution operator $U^\dagger$. We note that path probabilities of a dynamic Bayesian network can be determined experimentally \cite{mic21}. By now taking the ratio of Eqs.~\eqref{4} and \eqref{5}, and  inserting the microscopic first law \eqref{2}, we  obtain the  generalized non-Abelian detailed fluctuation relation 
\begin{equation}  \begin{aligned}
  \label{6}
    \frac{P(\Gamma)}{P^\dagger(\Gamma^\dagger)} =&  \exp{\boldsymbol{\lambda} \cdot ( \boldsymbol{w}+ \boldsymbol{\varepsilon})+ \Delta \boldsymbol{\lambda} \cdot \boldsymbol{q}}\\
    &\times\exp{-\Delta F^r - \Delta c - \Delta d},
  \end{aligned}
\end{equation}
where $\Delta \boldsymbol{\lambda}= \boldsymbol{\lambda}- \boldsymbol{\lambda}^R$ is the difference between system and bath affinities, and $\Delta F^r = F^{r,\tau} - F^{r,0}$ is the change of  free entropy, $ F^{r,t}= -\ln\Tr \big[\exp\big(-\boldsymbol{\lambda}^R \cdot \boldsymbol{A}^t\big) \big]$, associated with the  instantaneous generalized Gibbs   state $\pi^{r,t} = \exp \big(F^{r,t} -\boldsymbol{\lambda}^R\cdot \boldsymbol{A}^t  \big)$, which acts as a reference  state, with spectral decomposition $\pi^{r,t} =\sum_m p^{r,t}_m \Pi_m^{r,t}$.  This state generalizes the instantaneous equilibrium state in the standard derivation of the quantum fluctuation relation \cite{def11}. We have moreover introduced  the variation of the stochastic relative entropies of coherence \cite{str17}, $\Delta c(i,j,m,n) = \ln p^\tau_j/p^{d}_n - \ln p^0_i/p^{d}_m$,  and athermality \cite{bra13}, $\Delta d(m,n) = \ln p^{d}_n/p^{r}_n - \ln p^{d}_m/p^{r}_m$, of state $\pi$  with respect to state $\pi^{r}$ at the endpoints of the protocol; here $p^{d}$ refers to the probability of measuring the state $\pi$ in the eigenbasis of $\pi^{r}$.  The relative entropies of coherence and athermality respectively quantify the difference between the two states along the off-diagonal and diagonal \cite{str17,bra13}.

\begin{figure*}[t]
	\centering
	\begin{tikzpicture}
	\node (a) [label={[label distance=-.4 cm]145: \textbf{a)}}] at (0,0) {\includegraphics[width=0.32\textwidth]{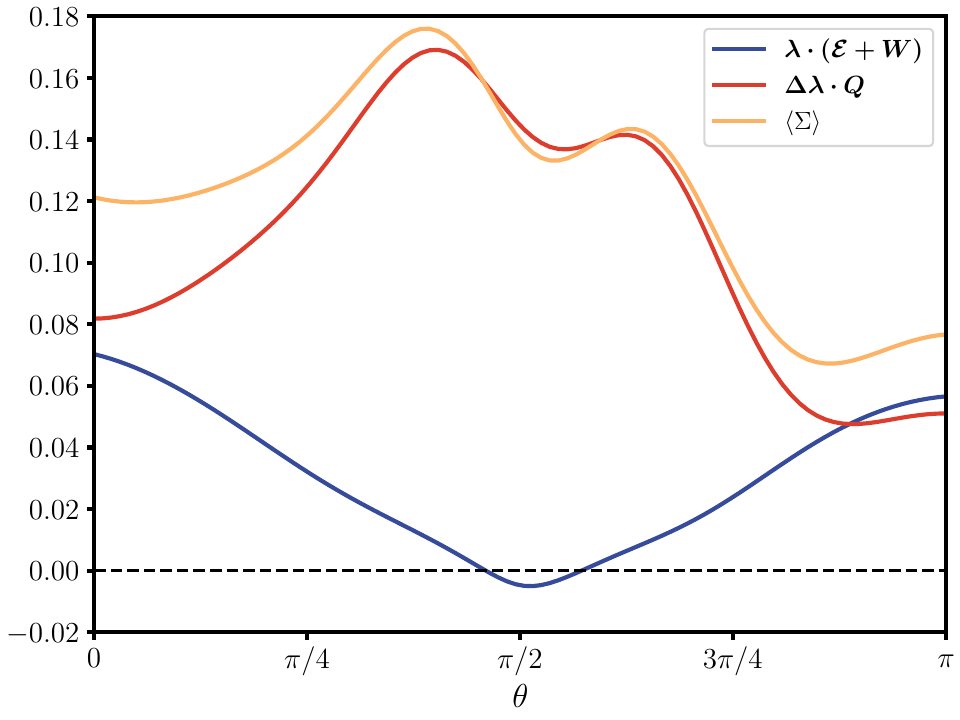}};	\node (a) [label={[label distance=-.4 cm]145: \textbf{b)}}] at (6,0) {\includegraphics[width=0.32\textwidth]{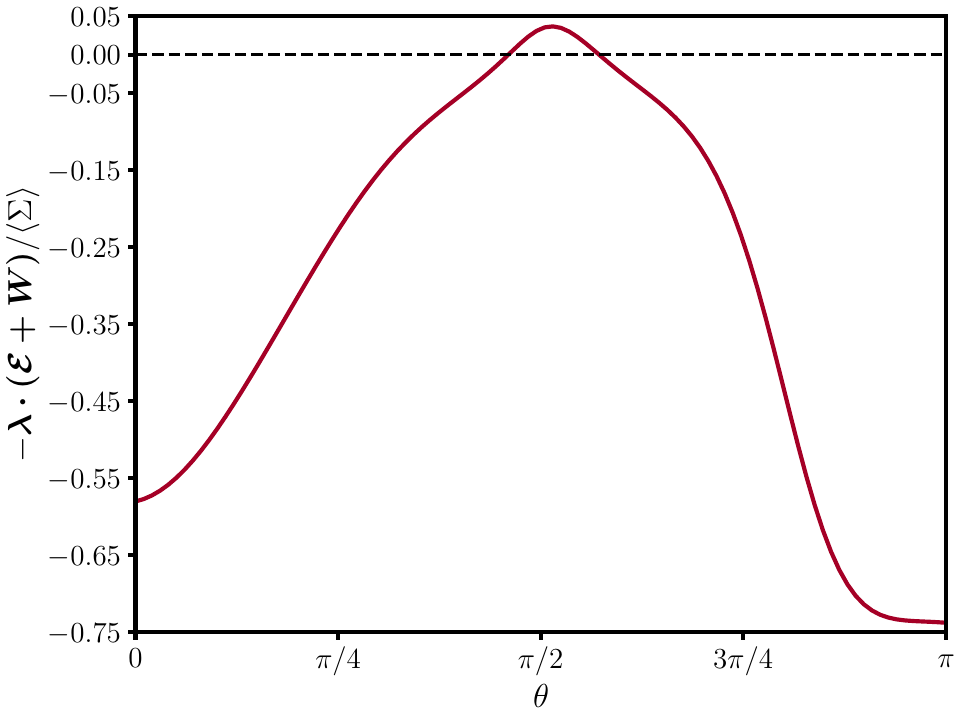}};
	\node (a) [label={[label distance=-.4 cm]145: \textbf{c)}}] at (12.0,0) {\includegraphics[width=0.32\textwidth]{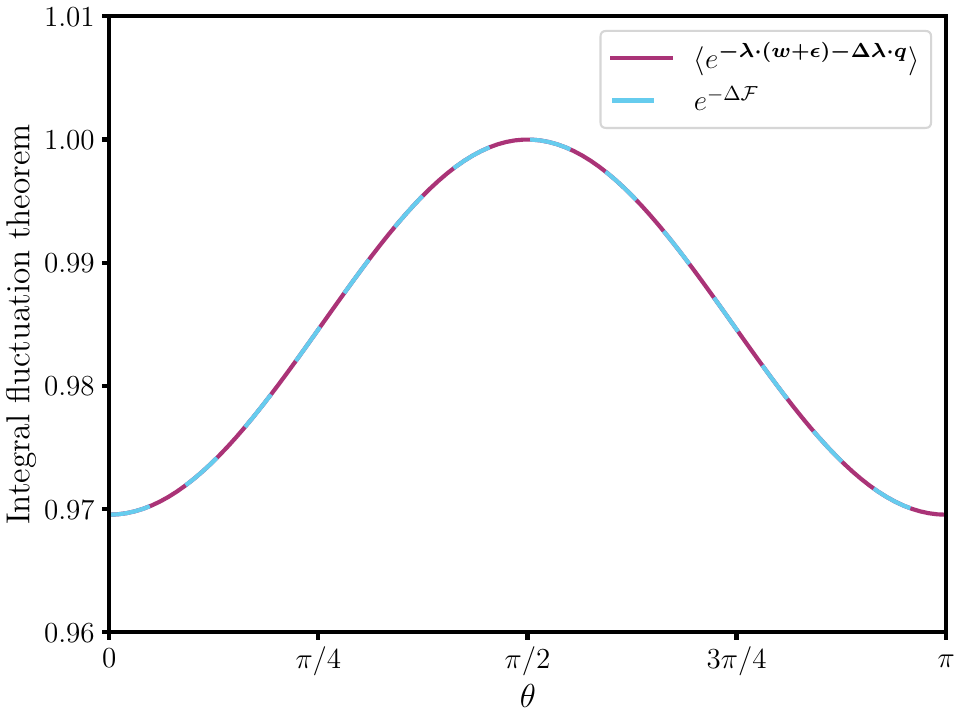}};
	\end{tikzpicture}
	\caption{Non-Abelian work fluctuation relation. a) Work $\boldsymbol{W}+ \boldsymbol{\mathcal{E}}$, heat $\boldsymbol{Q}$ and entropy production $\boldsymbol{\Sigma}$ for a linearly  driven system spin of the  two-qubit Heisenberg chain, as a function of the angle $\theta$ between the spin states. Work is done on the system in the Abelian case $\theta=0$. As coherences are increased, work is reduced before becoming negative around $\theta=\pi/2$. b) In this non-Abelian regime, work is extracted out of the system, and the ratio of extracted work and dissipation is maximum. c) The work fluctuation relation (7) is obeyed when the non-Abelian term $\boldsymbol{\mathcal{E}}$, Eq.~(3), is included (here $g_0 = 10$ and $g_\tau = 0.1$).}\label{fig2}
\end{figure*}

Equation \eqref{6} is an extension of the quantum Crooks relation for noncommuting conserved quantities. It represents  one of the most general nonequilibrium formulations of the quantum second law of thermodynamics   to date, incorporating not only the quantum properties of the states (such as discrete spectra) and of the dynamics (coherent time evolution), but also accounting for the first time for the presence of non-Abelian charges and induced quantum coherences.  It is important to realize that not including the non-Abelian term $\boldsymbol{\varepsilon}$ would lead to a violation of the fluctuation theorem, as well as of the first law \eqref{2}. The latter contribution is therefore essential for a correct analysis of the far-from-equilibrium thermodynamics of non-Abelian systems. Integrating Eq.~\eqref{6} over all forward trajectories further leads to the Jarzynski-like integral quantum fluctuation relation
\begin{equation}
\label{7}
\langle\exp{-\boldsymbol{\lambda} \cdot ( \boldsymbol{w}+ \boldsymbol{\varepsilon})- \Delta \boldsymbol{\lambda} \cdot \boldsymbol{q}+\Delta F^r + \Delta c + \Delta d}\rangle =1.
\end{equation}
Applying Jensen's inequality to Eq.~\eqref{7}, we additionally  have the generalized second-law like inequality
\begin{equation}\label{8}
  \boldsymbol{\lambda} \cdot ( \boldsymbol{W}+ \boldsymbol{\mathcal{E}})+ \Delta \boldsymbol{\lambda} \cdot \boldsymbol{Q}\geq \Delta F^r + \Delta \mathcal{C} + \Delta \mathcal{D},
\end{equation}
with the averaged thermodynamic variables $\boldsymbol{W} = \avg{\boldsymbol{w}}$, $\boldsymbol{\mathcal{E}} = \avg{\boldsymbol{\varepsilon}}$ and $\boldsymbol{Q} = \avg{\boldsymbol{q}}$, and the two information-theoretic quantities  $\Delta \mathcal{C} = \avg{\Delta c}$ and $\Delta \mathcal{D} = \avg{\Delta d}$ that correspond to the respective  difference in relative entropy of coherence and athermality between the two generalized Gibbs states. Equation \eqref{8} indicates that 
the amount of work extracted from the system is not only upper bounded  by the decrease of the  free entropy, as in standard commuting thermodynamics \cite{cal85},  but also by the non-Abelian work contribution $\boldsymbol{\mathcal{E}}$ and by the amount of coherence associated with the noncommutativity of the charges.  This (equilibrium) coherence is not dynamical, since it is solely specified by the difference  of the relative entropy of coherence between states $\pi_0$ and $\pi^r_0$ at the beginning of the protocol and the relative entropy of coherence between states $\pi_\tau$ and $\pi^r_\tau$ at the end of the process. As we will see in the numerical example below, these coherences will profoundly affect the amount of extractable work. We also note that the right-hand side of Eq.~\eqref{8} can be compactly  written as the difference of the nonequilibrium free entropy defined as $\mathcal{F} = F^r + \mathcal{C}+ \mathcal{D}$. 

When  system and environment have equal affinities,  $\boldsymbol{\lambda} = \boldsymbol{\lambda}^R$, then $\mathcal{F}$ reduces to $F$, and there is no heat exchange between the two since $\Delta \boldsymbol{ \lambda} = 0$. If the charges moreover commute, the non-Abelian work contribution vanishes, $\boldsymbol{\mathcal{E}} = 0$, and we  recover the inequality $ \boldsymbol{\lambda} \cdot \boldsymbol{W} \geq \Delta F$ obtained in Ref.~\cite{gur16} for generalized Gibbs ensembles with commuting conserved quantities.
Finally, following Ref.~\cite{lan21}, we may write  the average nonequilibrium entropy production, that characterizes the irreversible character of the quantum process, in the form 
\begin{equation}
 \label{9}
  \begin{split}
    \avg{\Sigma} =& \boldsymbol{\lambda} \cdot (\boldsymbol{W} + \boldsymbol{\mathcal{E}}) + \Delta \boldsymbol{ \lambda \cdot Q} - \Delta {\mathcal F} \geq 0.
  \end{split}
\end{equation}

\textit{Non-Abelian exchange fluctuation relation}. We proceed by examining the fluctuations of the exchange of charges  between two non-Abelian bodies with different generalized Gibbs states in the absence of external driving, and derive the corresponding transport fluctuation theorem. This situation can be viewed as a special case of the previous analysis in which  the system plays the role of one body  and the bath the role of the second one. 
Without mechanical driving, $ \boldsymbol{w}=0$, the first law \eqref{2} simplifies to 
$\Delta \boldsymbol{a}(i^0,j^\tau) =   \boldsymbol{q}(\mu,\nu) + \boldsymbol{\varepsilon}(i^0,\mu,j^\tau,\nu)$. Since there is no time dependence in the local charges of the system, the reference states are the same for the forward and backward protocols, implying that $\Delta \mathcal{F} = 0$. 
As a result, we have the non-Abelian exchange fluctuation relation
\begin{equation}  \begin{aligned}
  \label{10}
    \frac{P(\Gamma)}{P^\dagger(\Gamma^\dagger)} =&  \exp{\boldsymbol{\lambda} \cdot  \boldsymbol{\varepsilon}+ \Delta \boldsymbol{\lambda} \cdot \boldsymbol{q}},
  \end{aligned}
\end{equation}
as well as the corresponding integral version
\begin{equation}
\label{11}
\langle\exp{-\boldsymbol{\lambda} \cdot  \boldsymbol{\varepsilon}- \Delta \boldsymbol{\lambda} \cdot \boldsymbol{q}}\rangle =1.
\end{equation}
Equations \eqref{10} and \eqref{11} are extensions of the Jarzynski-Wojcik fluctuation theorem for heat transport \cite{jar04} valid for arbitrary non-Abelian states and associated charges. Applying again Jensen's inequality, we find in this case the mean nonequilibrium entropy production
\begin{equation}\label{12}
  \begin{split}
       \avg{\Sigma} =& \boldsymbol{\lambda} \cdot  \boldsymbol{\mathcal{E}} + \Delta \boldsymbol{ \lambda \cdot Q}  \geq 0.
  \end{split}
\end{equation}
The above expression reveals that  both charge transport and nonequilibrium entropy production are controlled by the non-Abelian work term  $\boldsymbol{\mathcal{E}}$.


\textit{Example of a  Heisenberg chain}. In order to gain deeper insight into the thermodynamics  of the non-Abelian term $\boldsymbol{\mathcal{E}}$, we examine a model of  two spins coupled via a Heisenberg interaction of the form $V = J \boldsymbol{\sigma} \cdot \boldsymbol{\sigma}^R$, where $\boldsymbol{\sigma}$ is a vector of Pauli matrices and $J$ is the coupling strength. This interaction preserves the average of $\sigma_i + \sigma^R_i$ for each Pauli matrix $\sigma_i$, thus constituting our charges. The local Hamiltonians of system and  environment are respectively taken as $H= \omega (\cos \theta \sigma_z + \sin \theta \sigma_x)/2$ and $H^R = \omega \sigma^R_z/2$, with frequency $\omega$. The corresponding initial states further read $\rho= \exp(F-\beta H)$ and $\rho^R = \exp(F^R -\beta^R H^R)$. The state $\rho$ is a non-Abelian generalized Gibbs state for $\theta \neq n\pi$ (with  $\lambda_x = \omega \sin \theta \beta / 2$ and $\lambda_z = \omega \cos \theta \beta / 2$), while  $\rho^R$ is a standard thermal state.

 We first consider the  case of transport between the undriven spins. Figure 1 shows the non-Abelian contribution ${\mathcal{E}}_z$, the  current $ {Q}_z$ and the entropy production $ \avg{\Sigma}_z$, as a function of the angle $\theta$  (for simplicity, we focus on the $z$-components). For $\theta=0$, the charges commute, the non-Abelian term vanishes and the  current is equal to the entropy production. We observe that ${\mathcal{E}}_z$ is negative when $\pi/2< \theta < \pi$ (Fig.~1a). In this noncommuting regime, $Q_z/\avg{\Sigma}_z$ is larger that one (Fig.~1b), revealing that the  current is enhanced (with respect to dissipation)  by the coherence
contained in the initial state  the basis of the charges. The exchange fluctuation relation (11) is only obeyed if $\boldsymbol{\mathcal{\varepsilon}}$ is included (Fig.~1c). We additionally drive the system via $H(t) = \omega [g(t) \cos \theta \sigma_z + \sin \theta \sigma_x]/2$, with the protocol $g(t) = g_0 (1 - t)/\tau + g_\tau t/\tau$  of duration $\tau$. In this instance, we see that work can only be extracted when $\theta$ is close to $\pi/2$, i.e., when the coherence of the initial state are as pronounced as possible (Fig.~2ab). The work fluctuation relation (7) is satisfied (Fig.~2c).

\textit{Conclusions}. Fluctuation theorems are essential to investigate the far-from-equilibrium thermodynamic properties of small quantum systems. We have derived generalizations of these relations, both in their work and exchange formulations, for generic non-Abelian thermal states with noncommuting conserved quantities, using the method of dynamic Bayesian networks. In doing so, we have determined a new non-Abelian contribution to both energy and entropy which is crucial to make correct thermodynamic predictions. By analyzing the example of a Heisenberg spin chain, we have further shown that this term can be tuned, and that it can be successfully exploited to enhance both work extraction and charge transport. This remarkable result indicates that non-Abelian states are a potential, currently untapped, physical resource for thermodynamic applications.

\textit{Acknowledgements.}
We would like to thank Kaonan Micadei, Thomas Jendrossek and Milton Aguilar for discussions, and acknowledge financial support from the German Research Foundation DFG (Grant FOR 2724). F. R. gratefully acknowledges the Georg H. Endress Foundation for financial support.

\textit{Note added.} While completing this manuscript, we became aware of a recent preprint (M. Scandi and G. Manzano, Universal statistics of charges exchanges in non-Abelian quantum transport, arXiv:2508.15540) presenting the derivation of the exchange fluctuation relations (10) and (11) using a collisional approach.

\clearpage

\begin{widetext}
\appendix

{\setcounter{page}{1} 

\section{Derivation of the non-Abelian contribution to the first law}

We here present the derivation of the non-Abelian term $\boldsymbol{\varepsilon}$, as given in Eq.~(3) of the main text.
We begin with the strict energy conservation condition,
\begin{equation}\label{charge_conservation}
  [V, A^t + A^R] = 0  
\end{equation}
at any point in the protocol. As the following calculation is true for every single charge, we choose to drop the bold letter for simplicity. We also assume, for the time being, that the Hamiltonian of the system is time independent (no mechanical driving). From Eq.~\eqref{charge_conservation}, we  then have 
\begin{equation}
  \begin{aligned}
    \mel{j^\tau, \nu}{[V, A^0]}{i^0, \mu} &= - \mel{j^\tau, \nu}{[V, A^R]}{i^0, \mu},\\
    \mel{j^\tau, \nu}{[V, A^\tau]}{i^0, \mu} &= - \mel{j^\tau, \nu}{[V, A^R]}{i^0, \mu},
  \end{aligned}
  \end{equation}
where we  took the matrix element between the initial and final measurements of the trajectory. We now sum both equations and divide by $2$ such that
\begin{equation}\label{charge_equality}
  \begin{aligned}
    \frac{1}{2}\left(\mel{j^\tau, \nu}{[V, A^0]}{i^0, \mu} + \mel{j^\tau, \nu}{[V, A^\tau]}{i^0, \mu}\right) = - \mel{j^\tau, \nu}{[V, A^R]}{i^0, \mu}.
  \end{aligned}
\end{equation}
We can further write the right-hand side as
\begin{equation}
  \begin{aligned}
    \mel{j^\tau, \nu}{[V, A^R]}{i^0, \mu} =& \mel{j^\tau, \nu}{V}{i^0, \mu} (\mel{\mu}{A^R}{\mu} - \mel{\nu}{A^R}{\nu})\\ 
    &+ \mel{j^\tau, \nu}{V (\mathds{1} - \Pi^R_\mu) A^R}{i^0, \mu} - \mel{j^\tau, \nu}{A^R (\mathds{1} - \Pi^R_\nu) V}{i^0, \mu}
  \end{aligned}
\end{equation}
by expanding the commutator and using the identity $\mathds{1} = \Pi^R_\alpha + (\mathds{1} - \Pi^R_\alpha)$, for $\alpha \in {\mu, \nu}$. Remembering that $q(\mu,\nu) = \mel{\mu}{A^R}{\mu} - \mel{\nu}{A^R}{\nu}$, we find
\begin{equation}\label{reservoir_commutator}
  \begin{aligned}
    \mel{j^\tau, \nu}{[V, A^R]}{i^0, \mu} =& \mel{j^\tau, \nu}{V}{i^0, \mu} q(\mu,\nu) + \mel{j^\tau, \nu}{V (\mathds{1} - \Pi^R_\mu) A^R}{i^0, \mu} - \mel{j^\tau, \nu}{A^R (\mathds{1} - \Pi^R_\nu) V}{i^0, \mu}.
  \end{aligned}
\end{equation}
A similar analysis can be done to the terms on the left-hand side of \cref{charge_equality}:
\begin{equation}
  \begin{aligned}
    \mel{j^\tau, \nu}{[V, A^0]}{i^0, \mu} &= \mel{j^\tau, \nu}{V}{i^0, \mu} \mel{i^0}{A^0}{i^0} + \mel{j^\tau, \nu}{V (\mathds{1} - \Pi^0_i) A^0}{i^0, \mu} - \mel{j^\tau, \nu}{A^0 V}{i^0, \mu},\\
    \mel{j^\tau, \nu}{[V, A^\tau]}{i^0, \mu} &= - \mel{j^\tau, \nu}{V}{i^0, \mu} \mel{j^\tau}{A^\tau}{j^\tau} - \mel{j^\tau, \nu}{A^\tau (\mathds{1} - \Pi^\tau_j) V}{i^0, \mu} + \mel{j^\tau, \nu}{V A^\tau}{i^0, \mu}.
  \end{aligned}
\end{equation}
We can now sum them and use $\Delta a(i^0,j^\tau) = \mel{j^\tau}{A^\tau}{j^\tau} - \mel{i^0}{A^0}{i^0}$ to yield
\begin{equation}
  \begin{aligned}
    \mel{j^\tau, \nu}{[V, A^0]}{i^0, \mu} + \mel{j^\tau, \nu}{[V, A^\tau]}{i^0, \mu} =& - \mel{j^\tau, \nu}{V}{i^0, \mu} \Delta a(i^0,j^\tau)\\
    &+\mel{j^\tau, \nu}{V (\mathds{1} - \Pi^0_i) A^0}{i^0, \mu} - \mel{j^\tau, \nu}{A^\tau(\mathds{1} - \Pi^\tau_j) V}{i^0, \mu}\\ 
    &+ \mel{j^\tau, \nu}{V A^\tau}{i^0, \mu}- \mel{j^\tau, \nu}{A^0 V}{i^0, \mu}.
  \end{aligned}
\end{equation}
If we sum and subtract $\Delta a(i^0,j^\tau)$, we can finally write
\begin{equation}\label{system_commutator}
  \begin{aligned}
    \mel{j^\tau, \nu}{[V, A^0]}{i^0, \mu} + \mel{j^\tau, \nu}{[V, A^\tau]}{i^0, \mu} =& - \mel{j^\tau, \nu}{V}{i^0, \mu} \left(2\Delta a(i^0,j^\tau) + \mel{j^\tau}{\Delta A}{j^\tau} - \mel{i^0}{\Delta A}{i^0}\right)\\
    &+\mel{j^\tau, \nu}{V \left(A^\tau (\mathds{1} - \Pi^0_i) + (\mathds{1} - \Pi^0_i) A^0 \right)}{i^0, \mu}\\
    &- \mel{j^\tau, \nu}{\left(A^\tau (\mathds{1} - \Pi^\tau_j) + (\mathds{1} - \Pi^\tau_j) A^0 \right)V}{i^0, \mu},
  \end{aligned}
\end{equation}
where $\Delta A = A^\tau - A^0$. Finally, we substitute \cref{reservoir_commutator,system_commutator} in \cref{charge_equality} to find the relation
\begin{equation}
  \begin{aligned}
    \Delta a(i^0,j^\tau) - q(\mu,\nu) = \varepsilon(i^0,j^\tau,\mu,\nu),
  \end{aligned}
\end{equation}
for the non-Abelian contribution
\begin{equation}
  \varepsilon(i^0,j^\tau,\mu,\nu) = \frac{\mel{j^\tau, \nu}{\mathcal{O}^\tau V - V \mathcal{O}^0}{i^0, \mu}}{\mel{j^\tau, \nu}{V}{i^0, \mu}}-\frac{\mel{j^\tau}{\Delta A}{j^\tau} - \mel{i^0}{\Delta A}{i^0}}{2},
\end{equation}
with
\begin{equation}
  \begin{aligned}
    \mathcal{O}^\tau &= \frac{A^\tau (\mathds{1} - \Pi^\tau_j) + (\mathds{1} - \Pi^\tau_j) A^0}{2} + A^R (\mathds{1} - \Pi^R_\nu),\\
    \mathcal{O}^0 &= \frac{A^\tau (\mathds{1} - \Pi^0_i) + (\mathds{1} - \Pi^0_i) A^0}{2} + (\mathds{1} - \Pi^R_\mu) A^R.
  \end{aligned}
\end{equation}
In the presence of external driving, the energy balance  equation needs to be extended to include the (mechanical) work done on the system.   We accordingly have the work performed by each charge,
\begin{equation}
  w(i^0,j^\tau,\mu,\nu) = \Delta a(i^0,j^\tau) - q(\mu,\nu) - \varepsilon(i^0,j^\tau,\mu,\nu),
\end{equation}
as is presented in the main text.
}
\end{widetext}

\begin{thebibliography}{0}%
\makeatletter
\providecommand \@ifxundefined [1]{%
 \@ifx{#1\undefined}
}%
\providecommand \@ifnum [1]{%
 \ifnum #1\expandafter \@firstoftwo
 \else \expandafter \@secondoftwo
 \fi
}%
\providecommand \@ifx [1]{%
 \ifx #1\expandafter \@firstoftwo
 \else \expandafter \@secondoftwo
 \fi
}%
\providecommand \natexlab [1]{#1}%
\providecommand \enquote  [1]{``#1''}%
\providecommand \bibnamefont  [1]{#1}%
\providecommand \bibfnamefont [1]{#1}%
\providecommand \citenamefont [1]{#1}%
\providecommand \href@noop [0]{\@secondoftwo}%
\providecommand \href [0]{\begingroup \@sanitize@url \@href}%
\providecommand \@href[1]{\@@startlink{#1}\@@href}%
\providecommand \@@href[1]{\endgroup#1\@@endlink}%
\providecommand \@sanitize@url [0]{\catcode `\\12\catcode `\$12\catcode `\&12\catcode `\#12\catcode `\^12\catcode `\_12\catcode `\%12\relax}%
\providecommand \@@startlink[1]{}%
\providecommand \@@endlink[0]{}%
\providecommand \url  [0]{\begingroup\@sanitize@url \@url }%
\providecommand \@url [1]{\endgroup\@href {#1}{\urlprefix }}%
\providecommand \urlprefix  [0]{URL }%
\providecommand \Eprint [0]{\href }%
\providecommand \doibase [0]{http://dx.doi.org/}%
\providecommand \selectlanguage [0]{\@gobble}%
\providecommand \bibinfo  [0]{\@secondoftwo}%
\providecommand \bibfield  [0]{\@secondoftwo}%
\providecommand \translation [1]{[#1]}%
\providecommand \BibitemOpen [0]{}%
\providecommand \bibitemStop [0]{}%
\providecommand \bibitemNoStop [0]{.\EOS\space}%
\providecommand \EOS [0]{\spacefactor3000\relax}%
\providecommand \BibitemShut  [1]{\csname bibitem#1\endcsname}%
\let\auto@bib@innerbib\@empty
\end{thebibliography}%


\begin{thebibliography}{99}
\bibitem{cal85} H. B. Callen, \textit{Thermodynamics and an Introduction to Thermostatistics}, (Wiley, New York, 1985).
\bibitem{jay57} E. T. Jaynes, Information Theory and Statistical Mechanics I, Phys. Rev. \textbf{106}, 620 (1957).
\bibitem{jay57a} E. T. Jaynes, Information Theory and Statistical Mechanics II, Phys. Rev. \textbf{108}, 171 (1957).
\bibitem{bal86} R. Balian, Y. Alhassid, and H. Reinhardt, A geometric approach based on information theory, Phys. Rep. \textbf{131}, 1 (1986).
\bibitem{bal87} R. Balian and N. Balazs, Equiprobability, inference and entropy in quantum theory, Ann. Phys.  N.Y.  \textbf{179}, 97  (1987). 

\bibitem{vid16} L. Vidmar and M. Rigol, Generalized Gibbs ensemble in integrable lattice models, J. Stat. Mech. \textbf{064007} (2016).
\bibitem{yun16} N. Yunger Halpern, P. Faist, J. Oppenheim, and A. Winter, Microcanonical and resource-theoretic derivations of the thermal state of a quantum system with noncommuting charges, Nat.
Commun. \textbf{7}, 12051 (2016).
\bibitem{gur16} Y. Guryanova, S. Popescu, A. J. Short, R. Silva, and P.
Skrzypczyk, Thermodynamics of quantum systems with multiple conserved quantities, Nat. Commun. \textbf{7}, 12049 (2016).
\bibitem{los17} M. Lostaglio, D. Jennings, and T. Rudolph, Thermodynamic resource theories, non-commutativity and maximum entropy principles, New J. Phys. \textbf{19},
043008 (2017).
\bibitem{pop20} S. Popescu, A. Belen Sainz, A. J. Short, and A. Winter, Reference Frames Which Separately Store Noncommuting Conserved Quantities, Phys. Rev. Lett. \textbf{125}, 090601 (2020).
\bibitem{fuk20} K. Fukai, Y. Nozawa, K. Kawahara, and T. N. Ikeda,
Noncommutative generalized Gibbs ensemble in isolated integrable quantum systems,
Phys. Rev. Research \textbf{2}, 033403 (2020).
\bibitem{yun20} N. Yunger Halpern, M. E. Beverland, and A. Kalev, Noncommuting conserved charges in quantum many-body thermalization
Phys. Rev. E \textbf{101}, 042117 (2020). 
\bibitem{man22} G. Manzano, J. M. R. Parrondo, and G. T. Landi, Non-Abelian Quantum Transport and Thermosqueezing Effects,
PRX Quantum \textbf{3}, 010304 (2022).
\bibitem{mad23} S. Majidy, W. F. Braasch, A. Lasek, T. Upadhyaya, A. Kalev,
and N. Y. Halpern, Noncommuting conserved charges in quantum thermodynamics and beyond, Nature Rev. Phys. \textbf{5}, 689 (2023).
\bibitem{upa24} T. Upadhyaya, W. F. Braasch, Jr., G. T. Landi, and N. Y.
Halpern, Non-Abelian Transport Distinguishes Three Usually
Equivalent Notions of Entropy Production, PRX Quantum \textbf{5}, 030355 (2024).
\bibitem{fra22} F. Kranzl, A. Lasek, M. K. Joshi, A. Kalev, R. Blatt, C. F. Roos, and N. Yunger Halpern, Experimental observation of thermalization with noncommuting charges,  PRX Quantum \textbf{4}, 020318 (2023).

\bibitem{esp09} M. Esposito, U. Harbola and S. Mukamel, Nonequilibrium fluctuations, fluctuation theorems, and counting statistics in quantum systems, Rev. Mod. Phys. \textbf{81}, 1665 (2009).
\bibitem{cam11} M. Campisi, P. H\"anggi, and P. Talkner,  Quantum Fluctuation Relations: Foundations and Applications, Rev. Mod. Phys., \textbf{83} 771 (2011).
\bibitem{def11} S. Deffner and E. Lutz, Nonequilibrium Entropy Production for Open Quantum Systems, Phys. Rev. Lett. \textbf{107}, 140404 (2011).
\bibitem{lan21} G. T. Landi and M. Paternostro, Irreversible entropy production: From classical to quantum, Rev. Mod. Phys. \textbf{93}, 035008 
(2021).
%

\bibitem{jar97} C. Jarzynski,  Nonequilibrium equality for free energy differences, Phys. Rev. Lett. \textbf{78}, 2690 (1997).
\bibitem{jar04} C. Jarzynski and D. K.  W\'ojcik,  Classical and quantum fluctuation theorems for heat exchange, Phys. Rev. Lett. \textbf{92}, 230602 (2004).
 
   
  \bibitem{hic14} J. M. Hickey and S. Genway, Fluctuation theorems and the generalized Gibbs ensemble in integrable systems, Phys. Rev. E \textbf{90}, 022107 (2014).
  \bibitem{mur18} J. Mur-Petit, A. Relano, R. A. Molina and D. Jaksch, Revealing missing charges
with generalised quantum fluctuation relations, Nat. Commun. \textbf{9}, 2006 (2018).
  \bibitem{mur20} J. Mur-Petit, A. Relano, R. A. Molina, and D. Jaksch, Fluctuations of work in realistic equilibrium states of quantum systems with conserved quantities, SciPost Phys. Proc. \textbf{3}, 024 (2020).
  
   
  \bibitem{str17} A. Streltsov, G. Adesso, and M. B. Plenio, Quantum coherence as a resource,  Rev.  Mod. Phys.  \textbf{89}, 041003 (2017).
  
  \bibitem{nea03} R. E. Neapolitan, \textit{Learning Bayesian Networks}, (Prentice Hall, Upper Saddle River, 2003).
\bibitem{dar09} A. Darwiche, \textit{Modeling and Reasoning with Bayesian Networks}, (Cambridge University Press, Cambridge, 2009).
 
\bibitem{mic20} K. Micadei, G. T. Landi, and E. Lutz
Quantum Fluctuation Theorems beyond Two-Point Measurements, Phys. Rev. Lett. \textbf{124}, 090602 (2020).
\bibitem{par20} J. J. Park, S. W. Kim, and V. Vedral, Fluctuation theorem for arbitrary quantum bipartite systems, Phys. Rev. E \textbf{101}, 052128 (2020).
\bibitem{str20} P. Strasberg, Thermodynamics of Quantum Causal Models: An Inclusive, Hamiltonian Approach, Quantum \textbf{4}, 240 (2020).
\bibitem{mic21} K. Micadei, J. P. S. Peterson, A. M. Souza, R. S. Sarthour, I. S. Oliveira, G. T. Landi, R. M. Serra, and E. Lutz, Experimental Validation of Fully Quantum Fluctuation Theorems Using Dynamic Bayesian Networks, Phys. Rev. Lett. \textbf{127}, 180603 (2021).
\bibitem{rod24} F. L. S. Rodrigues and E. Lutz, Nonequilibrium thermodynamics of quantum coherence beyond linear response, 
Comm. Phys. \textbf{7}, 61 (2024).
\bibitem{mic24} K. Micadei, G. T. Landi, and E. Lutz, Extracting Bayesian networks from multiple copies of a quantum system, EPL \textbf{144}, 60002 (2024).

\bibitem{bre02} H. P. Breuer and F. Petruccione, \textit{Open Quantum Systems}, (Oxford University Press, Oxford, 2002).
\bibitem{cic22} F. Ciccarello, S. Lorenzo, V. Giovannetti, and G. M. Palma, Quantum collision models: Open system dynamics from
repeated interactions, Phys. Rep. \textbf{954}, 1 (2022).

\bibitem{nie16} W. Niedenzu, D.  Gelbwaser-Klimovsky, A. G.  Kofman, and G.  Kurizki, On the
operation of machines powered by quantum non-thermal baths, New. J. Phys. \textbf{18}, 083012 (2016).
\bibitem{nie18} W. Niedenzu, V.  Mukherjee, A. Ghosh, A. G.  Kofman, and G.  Kurizki, Quantum engine efficiency bound beyond the second law of thermodynamics, Nat. Commun. \textbf{9}, 165 (2018).
\bibitem{man18} G. Manzano, Squeezed thermal reservoir as a generalized
equilibrium reservoir, Phys. Rev. E \textbf{98}, 042123 (2018).

\bibitem{tal07} P. Talkner, E. Lutz, and P. H\"anggi, Fluctuation theorems: Work is not an observab le, Phys. Rev. E \textbf{75}, 050102  (2007).

\bibitem{cro99} G. E. Crooks, Entropy production fluctuation theorem and the nonequilibrium work relation for free energy differences, Phys. Rev. E, \textbf{60} 2721 (1999).

\bibitem{bra13} G. S. L. Brand\~ao,  M.  Horodecki, J. Oppenheim, J. M.  Renes,  and R. W. Spekkens,  Resource Theory of Quantum States Out of Thermal Equilibrium, Phys. Rev. Lett. \textbf{111}, 250404  (2013).






\end{thebibliography}
\end{document}